\documentclass[a4paper]{report}
\usepackage[utf8]{inputenc}
\usepackage[T1]{fontenc}
\usepackage{RJournal}
\usepackage{amsmath,amssymb,array}
\usepackage{booktabs}

\providecommand{\tightlist}{%
  \setlength{\itemsep}{0pt}\setlength{\parskip}{0pt}}

\usepackage{mathtools} \usepackage[ruled, linesnumbered]{algorithm2e}

\begin{document}

\sectionhead{Contributed research article}
\volume{XX}
\volnumber{YY}
\year{20ZZ}
\month{AAAA}

\begin{article}

\title{Visual Diagnostics for Constrained Optimisation with Application
to Guided Tours}
\author{by H.Sherry Zhang, Dianne Cook, Ursula Laa, Nicolas
Langrené, and Patricia Menéndez}

\maketitle

\abstract{%
A guided tour helps to visualise high-dimensional data by showing
low-dimensional projections along a projection pursuit optimisation
path. Projection pursuit is a generalisation of principal component
analysis, in the sense that different indexes are used to define the
interestingness of the projected data. While much work has been done in
developing new indexes in the literature, less has been done on
understanding the optimisation. Index functions can be noisy, might have
multiple local maxima as well as an optimal maximum, and are constrained
to generate orthonormal projection frames, which complicates the
optimization. In addition, projection pursuit is primarily used for
exploratory data analysis, and finding the local maxima is also useful.
The guided tour is especially useful for exploration, because it
conducts geodesic interpolation connecting steps in the optimisation and
shows how the projected data changes as a maxima is approached. This
work provides new visual diagnostics for examining a choice of
optimisation procedure, based on the provision of a new data object
which collects information throughout the optimisation. It has helped to
diagnose and fix several problems with projection pursuit guided tour.
This work might be useful more broadly for diagnosing optimisers, and
comparing their performance. The diagnostics are implemented in the R
package \pkg{ferrn}.
}

\hypertarget{introduction}{%
\section{Introduction}\label{introduction}}

Visualisation is widely used in exploratory data analysis
\citep{tukey1977exploratory, unwin2015graphical, healy2018data, wilke2019fundamentals}.
Presenting information in graphics often unveils insights that would
otherwise not be discovered and provides a more comprehensive
understanding of the problem at hand. Task specific tools such as
\citet{li2020visualizing} show how visualisation can be used to
understand, for instance, the behaviour of the optimisation for the
example of neural network classification models. However, no general
visualisation tool is available for diagnosing optimisation procedures.
The work presented in this paper brings visualization tools into
optimisation problems with the aim to better understand the performance
of optimisers in practice.

The focus of this paper is on the optimisation problem arising in the
projection pursuit guided tour \citep{buja2005computational}, an
exploratory data analysis technique used for detecting interesting
structures in high-dimensional data through a set of lower-dimensional
projections \citep{cook2008grand}. The goal of the optimisation is to
identify the projection, represented by the projection matrix, that
gives the most interesting low-dimensional view. A view is said to be
interesting if it can show some structures of the data that depart from
normality, such as bimodality, clustering, or outliers.

The optimization challenges encountered in the projection pursuit guided
tour problem are common to those of optimization in general. Examples
include the existence of multiple optima (local and global), the
trade-off between computational burden and proximity to the optima, or
dealing with noisy objective functions that might be non-smooth and
non-differentiable \citep{jones1998efficient}. The visualization tools,
optimization methods and conceptual framework presented in this paper
can therefore be applied to other optimization problems.

The remainder of the paper is organised as follows. The next section
provides an overview of optimisation methods, specifically random search
and line search methods. A review of the projection pursuit guided tour,
an overview of the optimisation problem and outlines of three existing
algorithms follows. The third section presents the new visual
diagnostics, including the design of a data structure to capture
information during the optimisation, from which several diagnostic plots
are created. An illustration of how the diagnostic plots can be used to
examine the performance of different optimisers and guide improvements
to existing algorithms is shown using simulated data. Finally, an
explanation of the implementation in the R package, \CRANpkg{ferrn}
\citep{ferrn}, is provided.

\hypertarget{optim}{%
\section{Optimisation methods}\label{optim}}

The type of optimisation problem considered in this paper is constrained
optimization \citep{bertsekas2014constrained}, assuming it is not
possible to find a solution to the problem in the way of a closed-form.
That is, the problem consists in finding the minimum or maximum of a
function \(f \in L^p\) in the constrained space \(\mathcal{A}\).

Gradient-based methods are commonly used to optimise an objective
function, with the most notable one being the gradient ascent (descent)
method. Although these methods are popular, they rely on the
availability of the objective function derivatives. As will be shown in
the next section, the independent variables in our optimisation problem
are the entries of a projection matrix, and the computational time
required to perform differentiation on a matrix could impede the
rendering of tour animation. In addition, some objective functions rely
on the empirical distribution of the data, which makes it in general not
possible to get the gradient. Hence, gradient-based methods are not the
focus of this paper and consideration will be given to derivative-free
methods.

Derivative-free methods
\citep{conn2009introduction, rios2013derivative}, which do not rely on
the knowledge of the gradient, are more generally applicable.
Derivative-free methods have been developed over the years, where the
emphasis is on finding, in most cases, a near optimal solution. Here we
consider three derivative-free methods, two of which are random search
methods: creeping random search and simulated annealing, and the other
one is pseudo-derivative search.

Random search methods
\citep{Romeijn2009, zabinsky2013stochastic, andradottir2015review} have
a random sampling component as part of their algorithm and have been
shown to have the ability to optimise non-convex and non-smooth
functions. The initial random search algorithm, pure random search
\citep{Brooks1958-zk}, draws candidate points from the entire space
without using any information of the current position and updates the
current position when an improvement on the objective function is made.
As the dimension of the space becomes larger, sufficient sampling from
the entire space would require a long time for convergence to occur,
despite a guaranteed global convergence \citep{spall2005introduction}.
Various algorithms have thus been developed to improve pure random
search by either concentrating on a narrower sampling space or using a
different updating mechanism. Creeping random search
\citep{White1971-ci} is such a variation, where a candidate point is
generated within a neighbourhood of the current point. This makes
creeping random search faster to compute, however, global convergence is
no longer guaranteed. On the other hand, simulated annealing
\citep{kirkpatrick1983optimization, bertsimas1993simulated}, introduces
a different updating mechanism. Rather than only updating the current
point when an improvement is made, simulated annealing uses a Metropolis
acceptance criterion, where worse candidates still have a chance to be
accepted. The convergence of simulated annealing algorithms has been
widely researched \citep{mitra1986convergence, granville1994simulated}
and the global optimum can be attained under mild regularity conditions.

Pseudo-derivative search uses a common search scheme in optimisation:
line search. In line search methods, users are required to provide an
initial estimate \(x_{1}\) and, at each iteration, a search direction
\(S_k\) and a step size \(\alpha_k\) are generated. Then one moves on to
the next point following \(x_{k+1} = x_k + \alpha_kS_k\) and the process
is repeated until the desired convergence is reached. In derivative-free
methods, local information of the objective function is used to
determine the search direction. The choice of step size also needs
consideration, as inadequate step sizes might prevent the optimisation
method to converge to an optimum. An ideal step size can be chosen by
finding the value of \(\alpha_k \in \mathbb{R}\) that maximises
\(f(x_k + \alpha_kS_k)\) with respect to \(\alpha_k\) at each iteration.

\hypertarget{tour}{%
\section{Projection pursuit guided tour}\label{tour}}

Projection pursuit guided tour combines two different methods
(projection pursuit and guided tour) to explore interesting features in
a high-dimensional space. Projection pursuit, coined by
\citet{friedman1974projection}, detects interesting structures
(e.g.~clustering, outliers and skewness) in multivariate data via
low-dimensional projections. Guided tour \citep{cook1995grand} is one
variation of a broader class of data visualisation methods, tour
\citep{buja2005computational}, which displays high-dimensional data
through a series of animated projections.

Let \(\mathbf{X}_{n \times p}\) be the data matrix with \(n\)
observations in \(p\) dimensions. A \(d\)-dimensional projection is a
linear transformation from \(\mathbb{R}^p\) into \(\mathbb{R}^d\)
defined as \(\mathbf{Y} = \mathbf{X} \cdot \mathbf{A}\), where
\(\mathbf{Y}_{n \times d}\) is the projected data and
\(\mathbf{A}_{p\times d}\) is the projection matrix. We define
\(f: \mathbb{R}^{n \times d} \mapsto \mathbb{R}\) to be an index
function that maps the projected data \(\mathbf{Y}\) onto a scalar
value. This is commonly known as the projection pursuit index function,
or just index function, and is used to measure the ``interestingness''
of a given projection. An interesting projection shows structures that
are non-normal since theoretical proofs from
\citet{diaconis1984asymptotics} have shown that projections tend to be
normal, as \(n\) and \(p\) approach infinity under certain conditions.
There have been many index functions proposed in the literature, here
are a few examples: early indexes that can be categorised as measuring
the \(L^2\) distance between the projection and a normal distribution:
Legendre index \citep{friedman1974projection}, Hermite index
\citep{hall1989polynomial}, and natural Hermite index
\citep{cook1993projection}; chi-square index \citep{posse1995projection}
for detecting spiral structure; LDA index \citep{lee2005projection} and
PDA \citep{lee2010projection} index for supervised classification;
kurtosis index \citep{Loperfido2020} and skewness index
\citep{Loperfido2018} for detecting outliers in financial time series;
and most recently, scagnostic indexes \citep{laa2020using} for
summarising structures in scatterplot matrices based on eight scagnostic
measures.

As a general visualisation method, tour produces animations of
high-dimensional data via rotations of low-dimensional planes. There are
different versions depending on how the high-dimensional space is
investigated: grand tour \citep{cook2008grand} selects the planes
randomly to provide a general overview; manual tour
\citep{cook1997manual} gradually phases in and out one variable to
understand the contribution of that variable in the projection. Guided
tour, the main interest of this paper, chooses the planes with the aid
of projection pursuit, to gradually reveal the most interesting
projection. Given a random start, projection pursuit iteratively finds
bases with higher index values and the guided tour constructs a geodesic
interpolation between these planes to form a tour path. Figure
\ref{fig:tour-path} shows a sketch of the tour path where the blue
frames are produced by the projection pursuit optimisation and the white
frames are interpolations between the blue frames. Mathematical details
of the geodesic interpolation can be found in
\citet{buja2005computational}. The aforementioned tour method has been
implemented in the R package \CRANpkg{tourr} \citep{tourr}.

\begin{Schunk}
\begin{figure}

{\centering \includegraphics[width=0.7\linewidth]{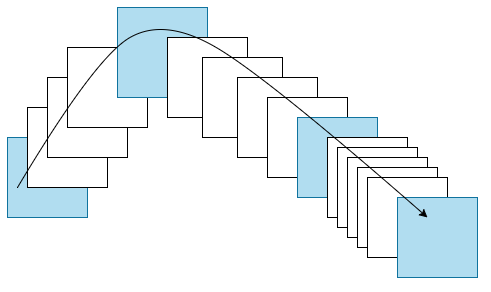}

}

\caption[An illustration for demonstrating the frames in a tour path]{An illustration for demonstrating the frames in a tour path. Each square (frame) represents the projected data with a corresponding basis. Blue frames are returned by the projection pursuit optimisation and white frames are constructed between two blue frames by geodesic interpolation.}\label{fig:tour-path}
\end{figure}
\end{Schunk}

\hypertarget{tour-optim}{%
\subsection{Optimisation in the tour}\label{tour-optim}}

In projection pursuit the optimisation aims at finding the global and
local maxima that give interesting projections according to an index
function. That is, it starts with a given randomly selected basis
\(\mathbf{A}_1\) and aims at finding an optimal final projection basis
\(\mathbf{A}_T\) that satisfies the following optimisation problem:

\begin{align}
\arg \max_{\mathbf{A} \in \mathcal{A}} f(\mathbf{X} \cdot \mathbf{A})  ~~~ s.t. ~~~ \mathbf{A}^{\prime} \mathbf{A} = I_d
\end{align}

\noindent where \(f\) and \(\mathbf{X}\) are defined as in the previous
section, \(\mathcal{A}\) is the set of all \(p\)-dimensional projection
bases, \(I_d\) is the \(d\)-dimensional identity matrix, and the
constraint ensures the projection bases, \(\mathbf{A}\), to be
orthonormal. It is worth noticing the following: 1) The optimisation is
constrained and the orthonormality constraint imposes a geometrical
structure on the bases space: it forms a Stiefel manifold. 2) There may
be index functions for which the objective function might not be
differentiable. 3) While finding the global optimum is the goal of the
optimisation problem, interesting projections may also appear in the
local optimum. 4) The optimisation should be fast to compute since the
tour animation is viewed by the users during the optimisation.

\hypertarget{existing-algorithms}{%
\subsection{Existing algorithms}\label{existing-algorithms}}

Three optimisers have been implemented in the \pkg{tourr} \citep{tourr}
package: creeping random search (CRS), simulated annealing (SA), and
pseudo-derivative (PD). Creeping random search (CRS) is a random search
optimiser that samples a candidate basis \(\mathbf{A}_{l}\) in the
neighbourhood of the current basis \(\mathbf{A}_{\text{cur}}\) by
\(\mathbf{A}_{l} = (1- \alpha)\mathbf{A}_{\text{cur}} + \alpha \mathbf{A}_{\text{rand}}\)
where \(\alpha \in [0,1]\) controls the radius of the sampling
neighbourhood and \(\mathbf{A}_{\text{rand}}\) is generated randomly.
\(\mathbf{A}_{l}\) is then orthonormalised to fulfil the basis
constraint. If \(\mathbf{A}_{l}\) has an index value higher than the
current basis \(\mathbf{A}_{\text{cur}}\), the optimiser outputs
\(\mathbf{A}_{l}\) for guided tour to construct an interpolation path.
The neighbourhood parameter \(\alpha\) is adjusted by a cooling
parameter: \(\alpha_{j+1} = \alpha_j * \text{cooling}\) before the next
iteration starts. The optimiser terminates when the maximum number of
iteration \(l_{\max}\) is reached before a better basis can be found.
The algorithm of CRS is summarised in Algorithm \ref{random-search} in
the appendix. \citet{posse1995projection} has proposed a slightly
different cooling scheme by introducing a halving parameter \(c\). In
his proposal \(\alpha\) is only adjusted if the last iteration takes
more than \(c\) times to find a better basis.

Simulated annealing (SA) uses the same sampling process as CRS but
allows a probabilistic acceptance of a basis with lower index value than
the current one. Given an initial value of \(T_0 \in \mathbb{R^{+}}\),
the ``temperature'' at iteration \(l\) is defined as
\(T(l) = \frac{T_0}{\log(l + 1)}\). When a candidate basis fails to have
an index value larger than the current basis, SA gives it a second
chance to be accepted with probability
\[P= \min\left\{\exp\left[-\frac{\mid I_{\text{cur}} - I_{l} \mid}{T(l)}\right],1\right\}\]
where \(I_{(\cdot)} \in \mathbb{R}\) denotes the index value of a given
basis. This implementation allows the optimiser to make a move and
explore the basis space even if the candidate basis does not have a
higher index value and hence enables the optimiser to jump out of a
local optimum. The algorithm \ref{simulated-annealing} in the appendix
highlights how SA differs from CRS in the inner loop.

Pseudo-derivative (PD) search uses a different strategy than CRS and SA.
Rather than randomly sample the basis space, PD first computes a search
direction by evaluating bases close to the current basis. A step size is
then chosen along the corresponding geodesic by another optimisation
over a 90 degree angle from \(-\pi/4\) to \(\pi/4\). The resulting
candidate basis \(\mathbf{A}_{**}\) is returned for the current
iteration if it has a higher index value than the current one. Algorithm
\ref{search-geodesic} in the appendix summarises the inner loop of the
PD.

\hypertarget{vis-diag}{%
\section{Visual diagnostics}\label{vis-diag}}

A data structure for diagnosing optimisers in projection pursuit guided
tour is first defined. With this data structure, four types of
diagnostic plots are presented.

\hypertarget{data-structure-for-diagnostics}{%
\subsection{Data structure for
diagnostics}\label{data-structure-for-diagnostics}}

Three main pieces of information are recorded during the projection
pursuit optimisation: 1) projection bases \(\mathbf{A}\), 2) index
values \(I\), and 3) state \(S\). For CRS and SA, possible states
include \texttt{random\_search}, \texttt{new\_basis}, and
\texttt{interpolation}. Pseudo-derivative (PD) has a wider variety of
states including \texttt{new\_basis}, \texttt{direction\_search},
\texttt{best\_direction\_search}, \texttt{best\_line\_search}, and
\texttt{interpolation}. Multiple iterators index the information
collected at different levels: \(t\) is a unique identifier prescribing
the natural ordering of each observation; \(j\) and \(l\) are the
counter of the outer and inner loop respectively. Other parameters of
interest recorded include \texttt{method} that tags the name of the
optimiser, and \texttt{alpha} that indicates the sampling neighbourhood
size for searching observations. A matrix notation of the data structure
is presented in equation \ref{eq:data-structure}.

\begin{equation}
\renewcommand\arraystretch{2}  
\begin{pmatrix}
t & \mathbf{A} & I & S & j &  l  & V_{1} & V_{2}\\
\hline
1 & \mathbf{A}_1 & I_1 & S_1 & 1 & 1 & V_{11} & V_{12}\\
\hline
2 & \mathbf{A}_2 & I_2 & S_2 & 2 & 1  & V_{21}  & V_{22}\\
\vdots & \vdots &\vdots &\vdots  &\vdots & \vdots &\vdots  &\vdots\\
\vdots & \vdots & \vdots &\vdots & 2 & l_2 & \vdots  & \vdots\\
\hline
\vdots &\vdots & \vdots &\vdots & 2  & 1& \vdots & \vdots\\
\vdots &\vdots &\vdots &\vdots &\vdots & \vdots & \vdots  &\vdots \\
\vdots &\vdots &\vdots &\vdots & 2 & k_2 &\vdots  & \vdots\\
\hline
\vdots &\vdots &\vdots &\vdots &\vdots & \vdots &\vdots &\vdots \\
\hline
\vdots & \vdots & \vdots &\vdots  & J &  1 & \vdots & \vdots \\
\vdots &\vdots &\vdots &\vdots &\vdots & \vdots &\vdots &\vdots \\
T & \mathbf{A}_T & I_T &S_T  & J &  l_{J} & V_{T1}& V_{T2}\\
\hline
\vdots &\vdots & \vdots &\vdots & J  & 1& \vdots & \vdots\\
\vdots &\vdots &\vdots &\vdots &\vdots & \vdots & \vdots  &\vdots \\
\vdots &\vdots &\vdots &\vdots & J & k_J &\vdots  & \vdots\\
\hline
\vdots& \vdots & \vdots & \vdots & J+1 & 1 & \vdots& \vdots\\
\vdots &\vdots &\vdots &\vdots &\vdots & \vdots &\vdots &\vdots \\
T^\prime & \mathbf{A}_{T^\prime} & I_{T^\prime} &S_{T^\prime}  & J+1 &  l_{J+1} & V_{T^\prime 1}& V_{T^\prime 2}\\
\end{pmatrix}
=
\begin{pmatrix}
\text{column name} \\
\hline
\text{search (start basis)} \\
\hline
\text{search} \\
\vdots \\
\text{search (accepted basis)} \\
\hline
\text{interpolate} \\
\vdots \\
\text{interpolate} \\
\hline
\vdots \\
\hline
\text{search} \\
\vdots \\
\text{search (final basis)} \\
\hline
\text{interpolate} \\
\vdots \\
\text{interpolate} \\
\hline
\text{search (no output)} \\
\vdots \\
\text{search (no output)} \\
\end{pmatrix}
\label{eq:data-structure}
\end{equation}

\noindent where \(T^{\prime} = T + k_{J}+ l_{J+1}\). Note that there is
no output in iteration \(J + 1\) since the optimiser does not find a
better basis in the last iteration and terminates. The final basis found
is \(A_T\) with index value \(I_T\).

The data structure constructed above meets the tidy data principle
\citep{wickham2014tidy} that requires each observation to form a row and
each variable to form a column. With tidy data structure, data wrangling
and visualisation can be significantly simplified by well-developed
packages such as \CRANpkg{dplyr} \citep{dplyr} and \CRANpkg{ggplot2}
\citep{ggplot2}.

\hypertarget{diagnostic-1-checking-how-hard-the-optimiser-is-working}{%
\subsection{Diagnostic 1: Checking how hard the optimiser is
working}\label{diagnostic-1-checking-how-hard-the-optimiser-is-working}}

A starting point of diagnosing an optimiser is to understand how many
searches it has conducted, i.e.~we want to summarise how the index is
increasing over iterations and how many basis need to be sampled at each
iteration. This is achieved using the function
\texttt{explore\_trace\_search()}: a boxplot shows the distribution of
index values for each try, where the accepted basis (corresponding to
the highest index value) is always shown as a point. In the case of few
tries at a given iteration, showing the data points directly may be
preferred over the boxplot, this is controlled via the \texttt{cutoff}
argument. Additional annotations are added to facilitate better reading
of the plot and these include 1) the number of points searched in each
iteration can be added as text label at the bottom of each iteration; 2)
the anchor bases to interpolate are connected and highlighted in a
larger size; and 3) the colour of the last iteration is in a greyscale
to indicate no better basis found in this iteration.

Figure \ref{fig:toy-search} shows an example of the search plot for CRS
(left) and SA (right). Both optimisers quickly find better bases in the
first few iterations and then take longer to find one in the later
iterations. The anchor bases, the ones found with the highest index
value in each iteration, always have an increased index value in the
optimiser CRS while this is not the case for SA. This feature gives CRS
an advantage in this simple example to quickly find the optimum.

\begin{Schunk}
\begin{figure}

{\centering \includegraphics[width=1\linewidth]{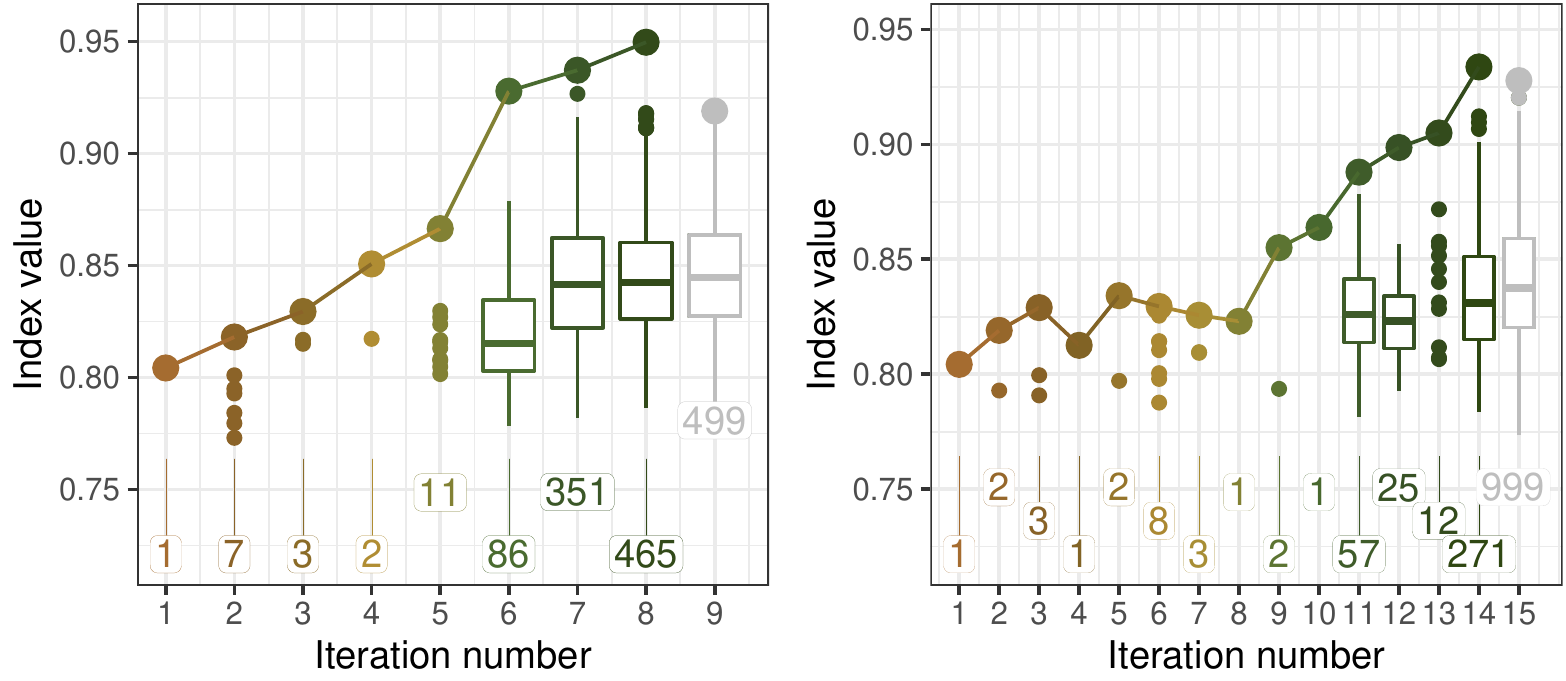}

}

\caption[A comparison of the searches by two optimisers]{A comparison of the searches by two optimisers: CRS (left) and SA (right) on a 2D projection problem of a six-variable dataset, \code{boa6} using holes index. Both optimisers reach the final basis with a similar index value while it takes SA longer to find the final basis. In the earlier iterations, optimisers quickly find a better basis to proceed while in the later iterations, most sampled bases fail to make an improvement on the index value and boxplot is used to summarise the distribution of the index values. There is no better basis found in the last iteration, 9 (left) and 15 (right), before reaching the maximum number of try and hence it is colored grey.}\label{fig:toy-search}
\end{figure}
\end{Schunk}

\hypertarget{toy-interp}{%
\subsection{Diagnostic 2: Examining the optimisation
progress}\label{toy-interp}}

\begin{Schunk}
\begin{figure}

{\centering \includegraphics[width=1\linewidth]{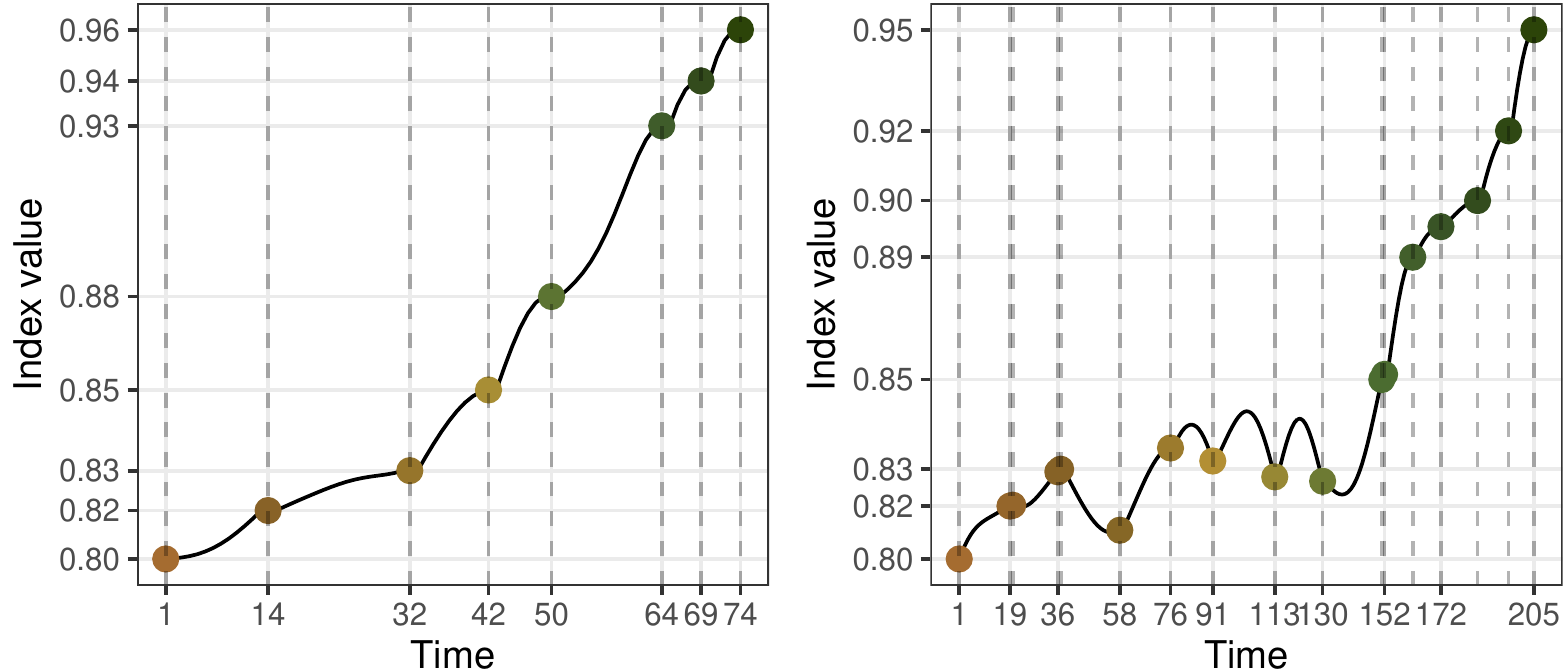}

}

\caption[An inspection of the index values as the optimisation progress for two optimisers]{An inspection of the index values as the optimisation progress for two optimisers: CRS (left) and SA (right). The holes index is optimised for a 2D projection problem on the six-variable dataset \code{boa6}. Lines indicate the interpolation and dots indicate new target bases generated by the optimisers. Interpolation in both optimisation is smooth while SA is observed to first pass by some bases with higher index value before reaching the target bases in time 76-130.}\label{fig:toy-interp}
\end{figure}
\end{Schunk}

Another interesting feature to examine is the changes in the index value
between interpolating bases since the projection on these bases is shown
in the tour animation. Trace plots are created by plotting the index
value against time. Figure \ref{fig:toy-interp} presents the trace plot
of the same optimisations as Figure \ref{fig:toy-search} and one can
observe that the trace is smooth in both cases. It may seem bizarre at
first sight that the interpolation sometimes passes bases with higher
index value before it decreases to a lower target. This happens because,
on the one hand, the probabilistic acceptance in SA implies that some
worse bases will be accepted by the optimiser. In addition, the guided
tour interpolates between the current and target basis to provide a
smooth transition between projections, and sometimes a higher index
value will be observed along the interpolation path. This indicates that
a non-monotonic interpolation cannot be avoided, even for CRS. Later, in
section \ref{monotonic}, there will be a discussion on improving the
non-monotonic interpolation for CRS.

\hypertarget{toy-pca}{%
\subsection{Diagnostic 3a: Understanding the optimiser's coverage of the
search space}\label{toy-pca}}

\begin{Schunk}
\begin{figure}

{\centering \includegraphics[width=1\linewidth]{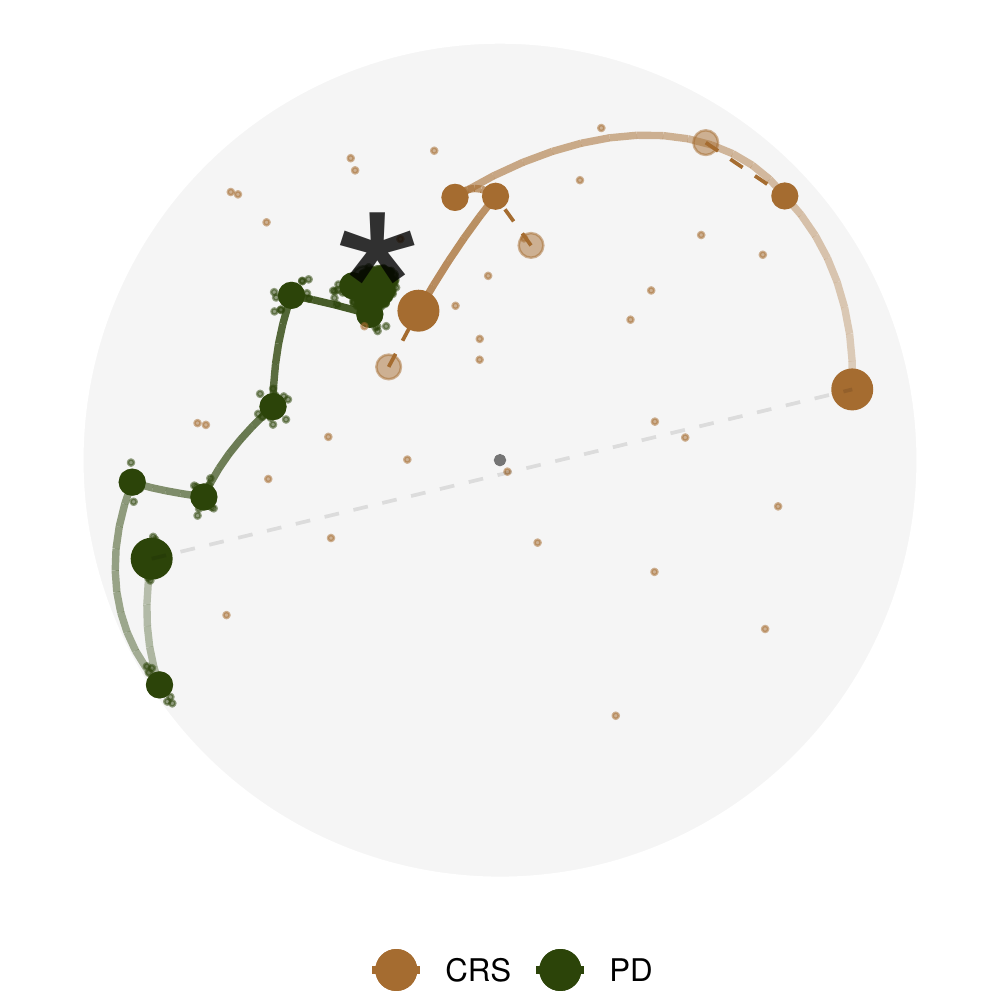}

}

\caption[Search paths of CRS (green) and PD (brown) in the PCA-reduced basis space for 1D projection problem on the five-variable dataset, \code{boa5} using holes index]{Search paths of CRS (green) and PD (brown) in the PCA-reduced basis space for 1D projection problem on the five-variable dataset, \code{boa5} using holes index. The basis space, a 5D unit sphere, is projected onto a 2D circle by PCA. All the bases in PD have been flipped for easier comparison of the final bases and a grey dashed line has been annotated to indicate the symmetry of the two start bases.}\label{fig:toy-pca}
\end{figure}
\end{Schunk}

Apart from checking the search and progression of an optimiser, looking
at where the bases are positioned in the basis space is also of
interest. Given the orthonormality constraint, the space of projection
bases \(\mathbf{A}_{p \times d}\) is a Stiefel manifold. For
one-dimensional projections, this forms a \(p\)-dimensional sphere. A
dimensionality reduction method, e.g.~principal component analysis is
applied to first project all the bases onto a 2D space. In a projection
pursuit guided tour optimisation, there are various types of bases
involved: 1) The starting basis; 2) The search bases that the optimiser
evaluated to produce the anchor bases; 3) The anchor bases that have the
highest index value in each iteration; 4) The interpolating bases on the
interpolation path; and finally, 5) the end basis. The importance of
these bases differs and the most important ones are the starting,
interpolating, and end bases. The anchor and search bases can be turned
on with argument \texttt{details\ =\ TRUE}. Sometimes, two optimisers
can start with the same basis but finish with bases of opposite sign.
This happens because the projection is invariant to the orientation of
the basis and so is the index value. However, this creates difficulties
for comparing the optimisers since the end bases will be symmetric to
the origin. A sign flipping step is conducted to flip the signs of all
the bases in one routine if different optimisations finish at opposite
places.

Several annotations have been made to help understanding this plot. As
mentioned previously, the original basis space is a high-dimensional
sphere and random bases on the sphere can be generated via the
\CRANpkg{geozoo} \citep{geozoo} package. Along with the bases recorded
during the optimisation and a zero basis, the parameters (centre and
radius) of the 2D space can be estimated by PCA. The centre of the
circle is the first two PC coordinates of the zero matrix, and the
radius is estimated as the largest distance from the centre to the
basis. The theoretical best basis is known for simulated data and can be
labelled to compare how close the end basis found by the optimisers is
to the theoretical best. Various aesthetics, i.e.~size, alpha
(transparency), and colour, are applicable to emphasize critical
elements and adjust for the presentation. For example, anchor points and
search points are less important and hence a smaller size and alpha are
used. Alpha can also be applied on the interpolation paths to show the
start to finish from transparent to opaque.

Figure \ref{fig:toy-pca} shows the PCA plot of CRS and PD for a 1D
projection problem. Both optimisers find the optimum but PD gets closer.
With the PCA plot, one can visually appreciate the nature of these two
optimisers: PD first evaluates points in a small neighbourhood for a
promising direction, while CRS evaluates points randomly in the search
space to search for the next target. There are dashed lines annotated
for CRS and it describes the interruption of the interpolation, which
will be discussed in detail in Section \ref{monotonic}.

\hypertarget{diagnostic-3b-animating-the-diagnostic-plots}{%
\subsection{Diagnostic 3b: Animating the diagnostic
plots}\label{diagnostic-3b-animating-the-diagnostic-plots}}

\begin{Schunk}
\begin{figure}

{\centering \includegraphics[width=1\linewidth]{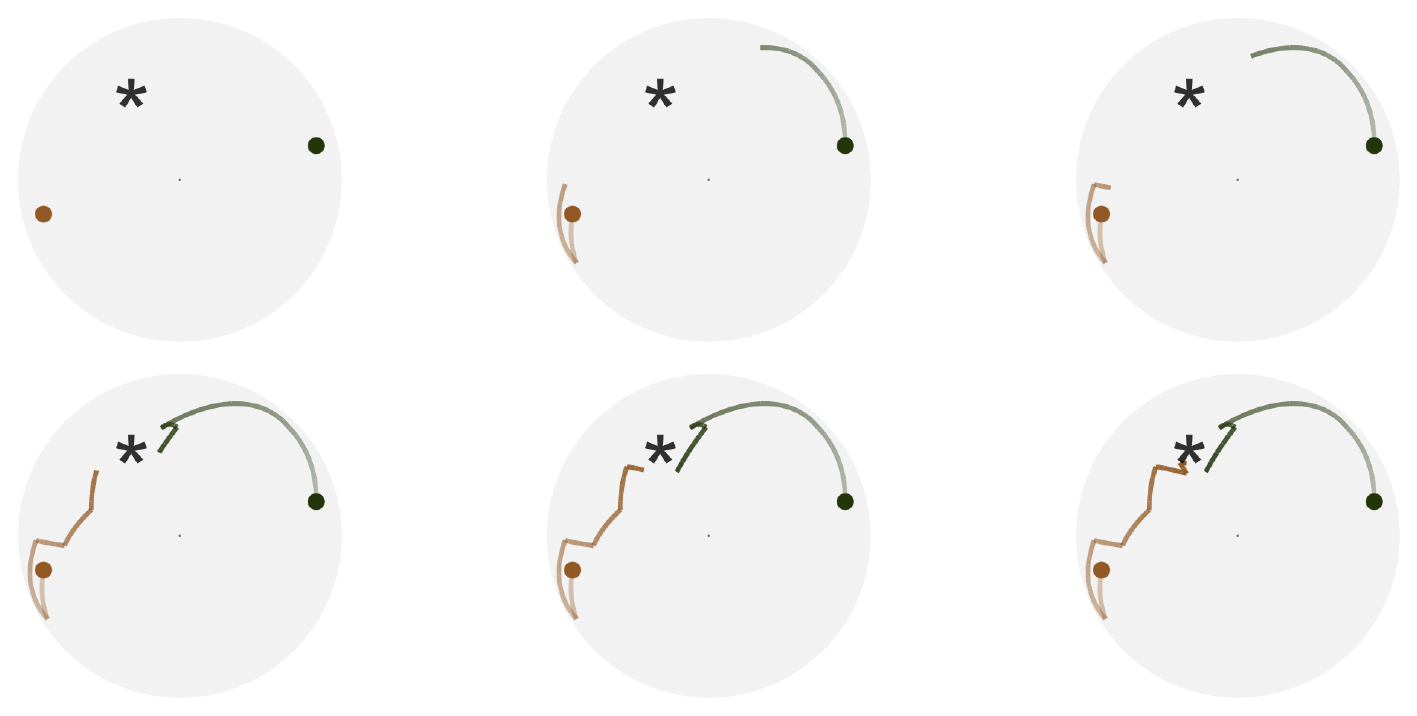}

}

\caption{Six frames selected from the animated version of the previous plot. With animation, the progression of the search paths from start to finish is better identified. CRS (green) finishes the optimisation quicker than PD (brown) since there is no further movement for CRS in the sixth frame. The full video of the animation can be found at \url{https://vimeo.com/504242845}.}\label{fig:toy-pca-animated}
\end{figure}
\end{Schunk}

An animation is another type of display to show how the search
progresses from start to finish in the space. An
\texttt{animate\ =\ TRUE} argument is used to enable the animation in
the PCA plot. Figure \ref{fig:toy-pca-animated} shows six frames from
the animation of the PCA plot in Figure \ref{fig:toy-pca}. An additional
piece of information one can learn from this animation is that CRS finds
its end basis quicker than PD since CRS finishes its search in the 5th
frame while PD is still making more progress.

\hypertarget{diagnostic-4a-the-tour-looking-at-itself}{%
\subsection{Diagnostic 4a: The tour looking at
itself}\label{diagnostic-4a-the-tour-looking-at-itself}}

\begin{Schunk}
\begin{figure}

{\centering \includegraphics[width=1\linewidth]{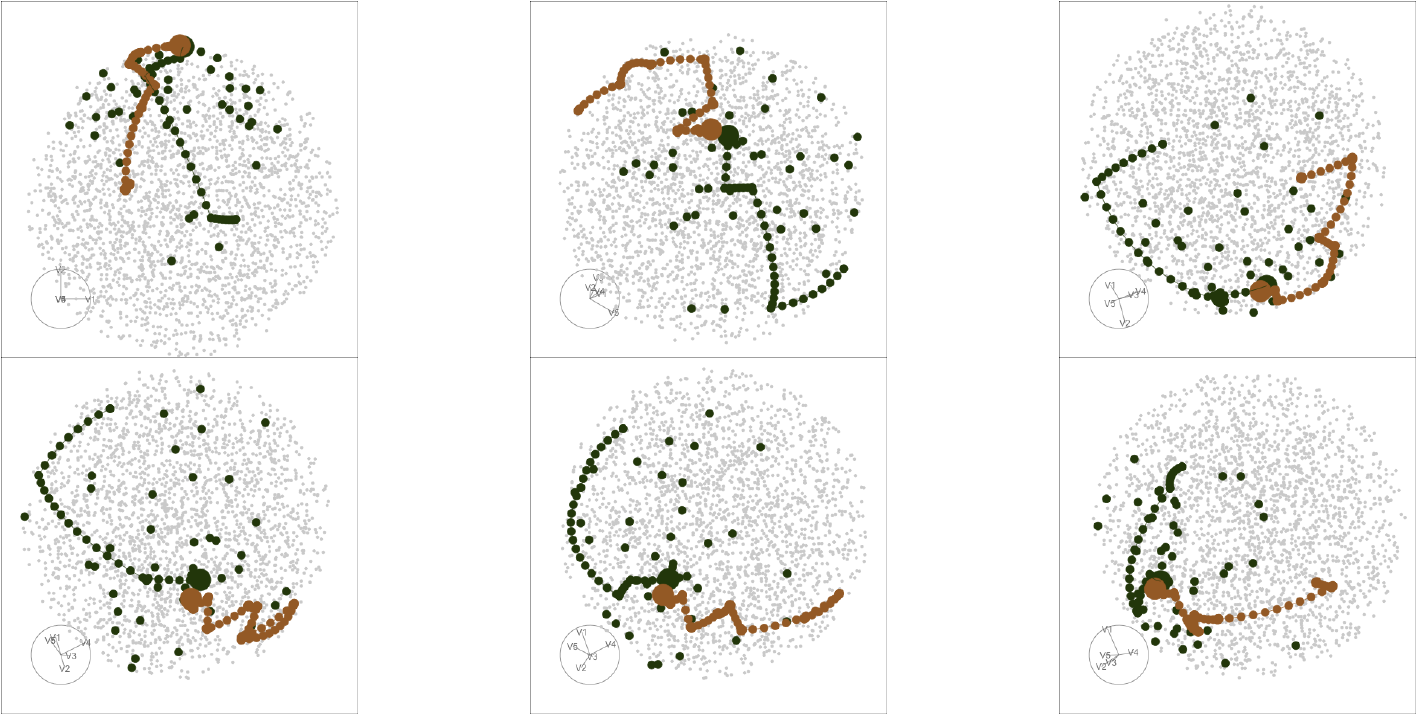}

}

\caption{Six frames selected from rotating the high-dimensional basis space, along with the same two search paths from Figure \ref{fig:toy-pca} and \ref{fig:toy-pca-animated}. The basis space in this example is a 5D unit sphere, on which points (grey) are randomly generated via the CRAN package \CRANpkg{geozoo}. The full video of the animation can be found at \url{https://vimeo.com/512885379}.}\label{fig:toy-tour}
\end{figure}
\end{Schunk}

As mentioned previously, the original \(p \times d\) dimension space can
be simulated via randomly generated bases in the \pkg{geozoo}
\citep{geozoo} package. While the PCA plot projects the bases from the
direction that maximises the variance, the tour plot displays the
original high-dimensional space from various directions using animation.
Figure \ref{fig:toy-tour} shows some frames from the tour plot of the
same two optimisations in its original 5D space.

\hypertarget{diagnostic-4b-forming-a-torus}{%
\subsection{Diagnostic 4b: Forming a
torus}\label{diagnostic-4b-forming-a-torus}}

While the previous few examples have looked at the space of 1D bases in
a unit sphere, this section visualises the space of 2D bases. Recall
that the columns in a 2D basis are orthogonal to each other, so the
space of \(p \times 2\) bases is a \(p\)-D torus. For \(p = 3\) one
would see a classical 3D torus shape as shown by the grey points in
Figure \ref{fig:toy-torus}. The two circles of the torus can be observed
to be perpendicular to each other and this can be linked back to the
orthogonality condition. Two paths from CRS and PD are plotted on top of
the torus and coloured in green and brown respectively to match the
previous plots. The final basis found by PD and CRS are shown in a
larger shape and printed below respectively:

\begin{Schunk}
\begin{Soutput}
#>              [,1]        [,2]
#> [1,]  0.001196285  0.03273881
#> [2,] -0.242432715  0.96965761
#> [3,] -0.970167484 -0.24226493
\end{Soutput}
\end{Schunk}

\begin{Schunk}
\begin{Soutput}
#>             [,1]         [,2]
#> [1,]  0.05707994 -0.007220138
#> [2,] -0.40196202 -0.915510160
#> [3,] -0.91387549  0.402230054
\end{Soutput}
\end{Schunk}

Both optimisers have found the third variable in the first direction,
and the second variable in the second direction. Note however the
different orientation of the basis, following from the different sign in
the second column. One would expect to see this in the torus plot as the
final bases match each other when projected onto one torus circle (due
to the same sign in the first column) and are symmetric when projected
onto the other (due the sign difference in the second column). In Figure
\ref{fig:toy-torus}, this can be seen most clearly in frame 5 where the
two circles are rotated into a line from our view.

\begin{Schunk}
\begin{figure}

{\centering \includegraphics[width=1\linewidth]{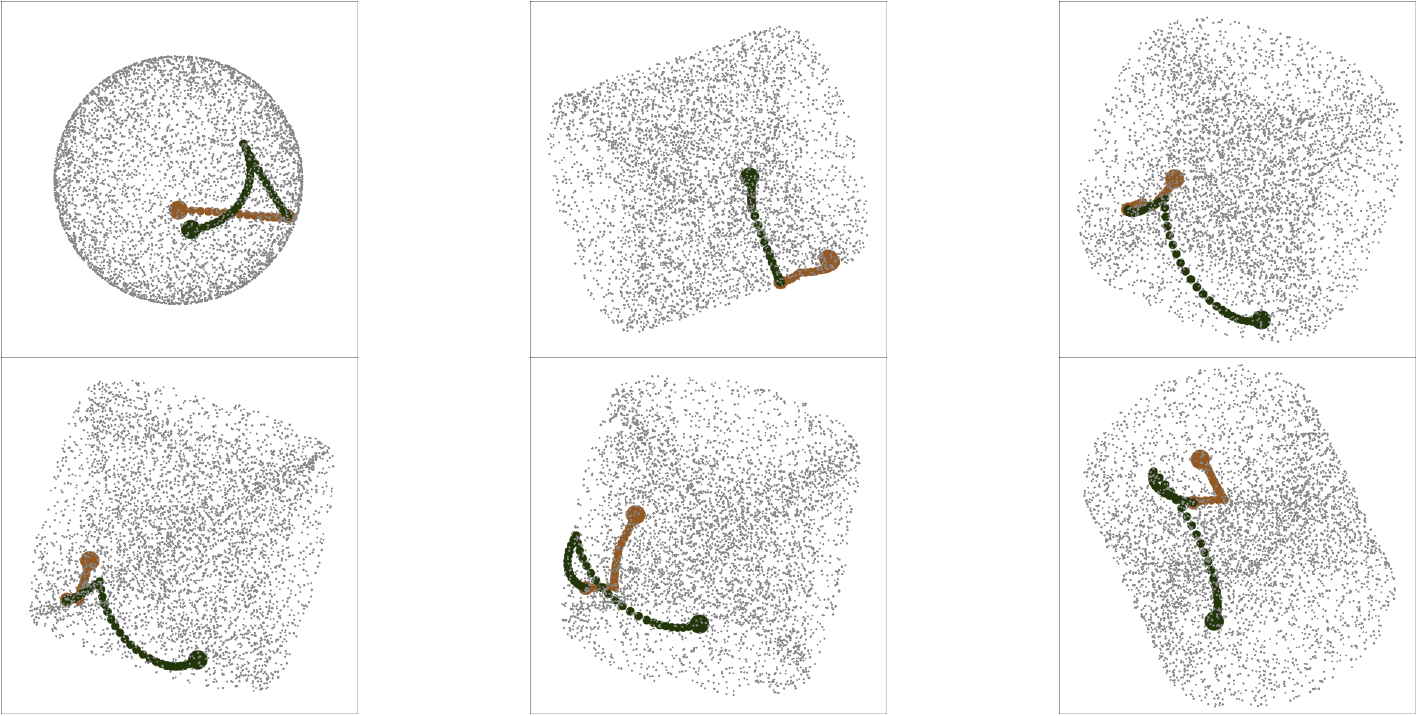}

}

\caption{Six frames selected from rotating the 2D basis space along with two search paths optimised by PD (brown) and CRS (green). The projection problem is a 2D projection with three variables using the holes index. The grey points are randomly generated 2D projection bases in the space and it can be observed that these points form a torus. The full video of the animation can be found at \url{https://vimeo.com/512882784}.}\label{fig:toy-torus}
\end{figure}
\end{Schunk}

\hypertarget{application}{%
\section{Diagnosing an optimiser}\label{application}}

In this section, several examples will be presented to show how the
diagnostic plots discover something unexpected in projection pursuit
optimisation, and guide the implementation of new features.

\hypertarget{simulation-setup}{%
\subsection{Simulation setup}\label{simulation-setup}}

Random variables with different distributions have been simulated and
the distributional form of each variable is presented in equations
\ref{eq:sim-norm} to \ref{eq:sim-x7}. Variables \texttt{x1}, \texttt{x8}
to \texttt{x10} are normal noise with zero mean and unit variance and
\texttt{x2} to \texttt{x7} are normal mixtures with varied weights and
locations. All the variables have been scaled to have an overall unit
variance before projection pursuit. The holes index
\citep{cook2008grand}, used for detecting bimodality of the variables,
is used throughout the examples unless otherwise specified.

\begin{align}
x_1 \overset{d}{=} x_8 \overset{d}{=} x_9 \overset{d}{=} x_{10}& \sim \mathcal{N}(0, 1) \label{eq:sim-norm} \\
x_2 &\sim 0.5 \mathcal{N}(-3, 1) + 0.5 \mathcal{N}(3, 1)\label{eq:sim-x2}\\
\Pr(x_3) &=
\begin{cases}
0.5 & \text{if $x_3 = -1$ or $1$}\\
0 & \text{otherwise}
\end{cases}\label{eq:sim-x3}\\
x_4 &\sim 0.25 \mathcal{N}(-3, 1) + 0.75 \mathcal{N}(3, 1) \label{eq:sim-x4}\\
x_5 &\sim \frac{1}{3} \mathcal{N}(-5, 1) + \frac{1}{3} \mathcal{N}(0, 1) + \frac{1}{3} \mathcal{N}(5, 1)\label{eq:sim-x5}\\
x_6 &\sim 0.45 \mathcal{N}(-5, 1) + 0.1 \mathcal{N}(0, 1) + 0.45 \mathcal{N}(5, 1)\label{eq:sim-x6}\\
x_7 &\sim 0.5 \mathcal{N}(-5, 1) + 0.5 \mathcal{N}(5, 1)
\label{eq:sim-x7}
\end{align}

\hypertarget{monotonic}{%
\subsection{A problem of non-monotonicity}\label{monotonic}}

An example of non-monotonic interpolation has been given in Figure
\ref{fig:toy-interp}: a path that passes bases with higher index value
than the target one. For SA, a non-monotonic interpolation is justified
since target bases do not necessarily have a higher index value than the
current one, while this is not the case for CRS. The original trace plot
for a 2D projection problem, optimised by CRS, is shown on the left
panel of Figure \ref{fig:interruption} and one can observe that the
non-monotonic interpolation has undermined the optimiser to realise its
full potential. Hence, an interruption is implemented to stop at the
best basis found in the interpolation. The right panel of Figure
\ref{fig:interruption} shows the trace plot after implementing the
interruption and while the first two interpolations are identical, the
basis at time 61 has a higher index value than the target in the third
interpolation. Rather than starting the next iteration from the target
basis on time 65, CRS starts the next iteration at time 61 on the right
panel and reaches a better final basis.

\begin{Schunk}
\begin{figure}

{\centering \includegraphics[width=1\linewidth]{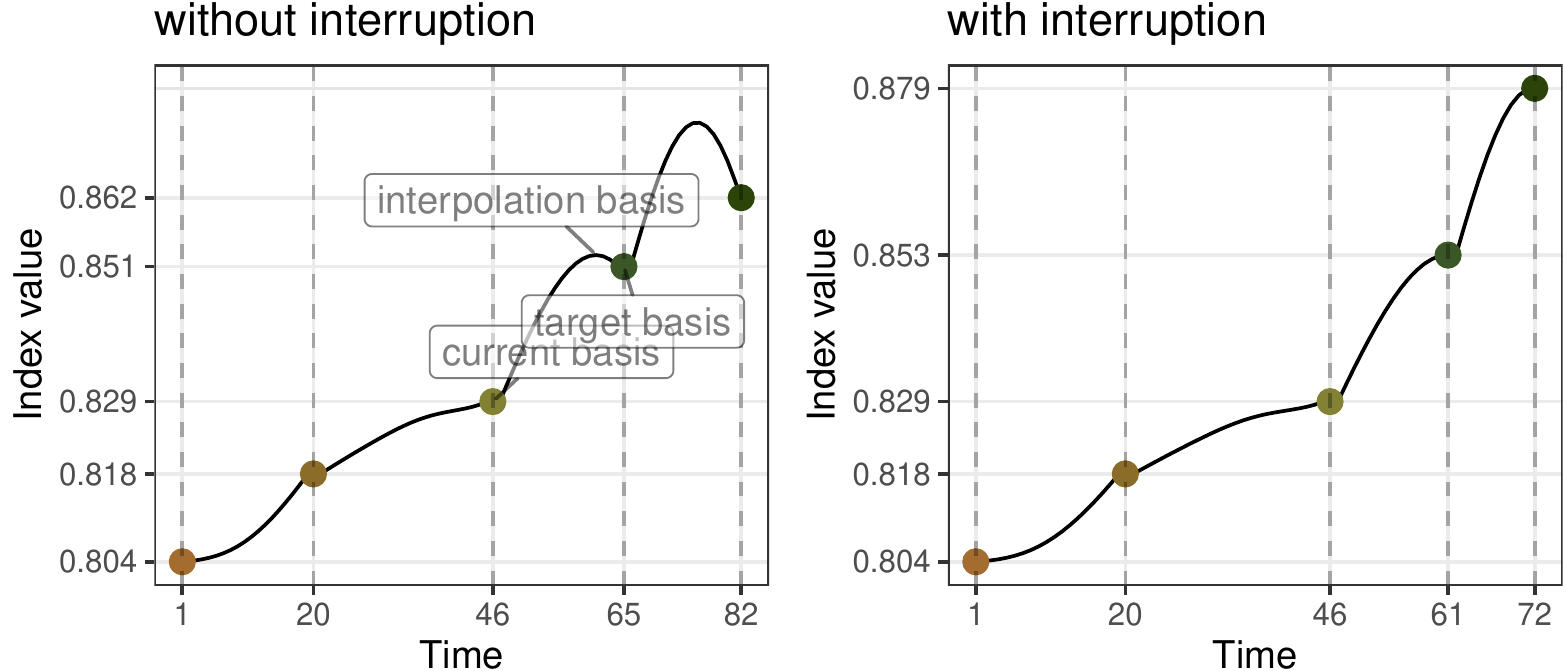}

}

\caption[Comparison of the interpolation before and after implementing the interruption for the 2D projection problem on \code{boa6} data using holes index, optimised by CRS]{Comparison of the interpolation before and after implementing the interruption for the 2D projection problem on \code{boa6} data using holes index, optimised by CRS. On the left panel, basis with higher index value is found during the interpolation but not used. On the right panel, the interruption stops the interpolation at the basis with the highest index value for each iteration and results in a final basis with higher index value, as shown on the right panel.}\label{fig:interruption}
\end{figure}
\end{Schunk}

\hypertarget{close-but-not-close-enough}{%
\subsection{Close but not close
enough}\label{close-but-not-close-enough}}

\begin{Schunk}
\begin{figure}

{\centering \includegraphics[width=1\linewidth]{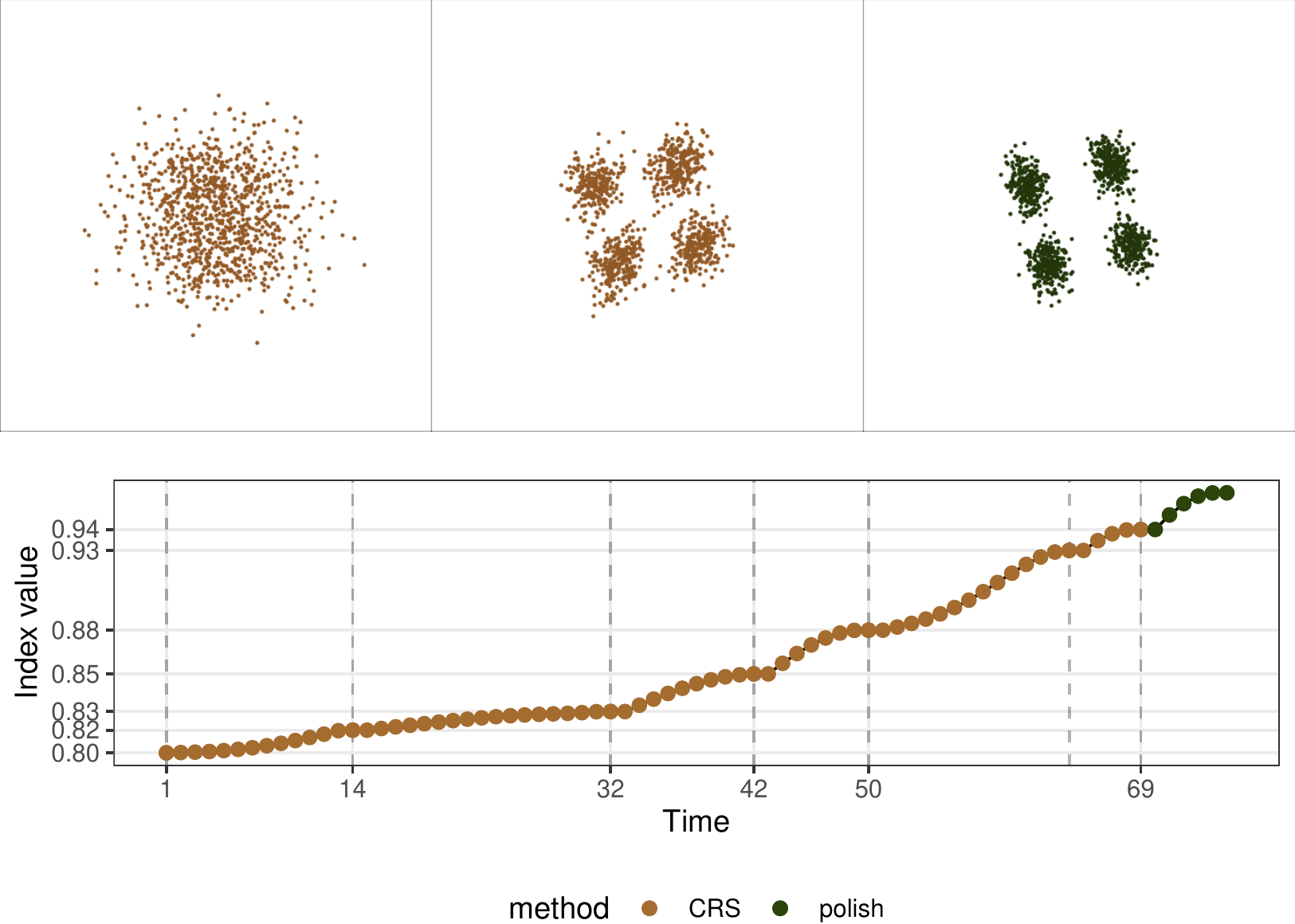}

}

\caption[Comparison of the projected data before and after using polishing for a 2D projection problem on \code{boa6} data using holes index]{Comparison of the projected data before and after using polishing for a 2D projection problem on \code{boa6} data using holes index. The top row shows the initial projected data and the final views after CRS and polish search and the second row traces the index value. The clustering structure in the data is detected by CRS (top middle panel) but the polish step improves the index value and produces clearer boundaries of the clusters (top right panel), especially along the vertical axis.}\label{fig:polish}
\end{figure}
\end{Schunk}

Once the final basis has been found by an optimiser, one may want to
push further in the close neighbourhood to see if an even better basis
can be found. A polish search takes the final basis of an optimiser as
the start of a new guided tour to search for local improvements. The
polish algorithm is similar to the CRS but with three distinctions: 1) a
hundred rather than one candidate bases are generated each time in the
inner loop; 2) the neighbourhood size is reduced in the inner loop,
rather than in the outer loop; and 3) three more termination conditions
have been added to ensure the new basis generated is distinguishable
from the current one in terms of the distance in the space, the relative
change in the index value, and neighbourhood size:

\begin{enumerate}
\def\labelenumi{\arabic{enumi})}
\tightlist
\item
  the distance between the basis found and the current basis needs to be
  larger than 1e-3;
\item
  the relative change of the index value needs to be larger than 1e-5;
  and
\item
  the alpha parameter needs to be larger than 0.01
\end{enumerate}

Figure \ref{fig:polish} presents the projected data and trace plot of a
2D projection, optimised by CRS and followed by the polish step. The top
row shows the initial projection, the final projection after CRS, and
the final projection after polish, respectively The end basis found by
CRS reveals the four clusters in the data, but the edges of each cluster
are not clean-cut. Polish works with this end basis and further pushes
the index value to produce clearer edges of the cluster, especially
along the vertical axis.

\hypertarget{seeing-the-signal-in-the-noise}{%
\subsection{Seeing the signal in the
noise}\label{seeing-the-signal-in-the-noise}}

The holes index function used for all the examples before this section
produces a smooth interpolation, while this is not the case for all the
indexes An example of a noisy index function for 1D projections compares
the projected data, \(\mathbf{Y}_{n \times 1}\), to a randomly generated
normal distribution, \(\mathcal{N}_{n \times 1}\), using the Kolmogorov
test. Let \(F_{.}(n)\) be the ECDF function (empirical cummulative
distribution function), with two possible subscripts \(Y\) and
\(\mathcal{N}\) representing the projected and randomly generated data,
and \(n\) denoting the number of observation, the Normal Kolmogorov
index \(I^{nk}(n)\), is defined as:

\[I^{K}(n) = \max \left[F_{Y}(n) - F_{\mathcal{N}}(n)\right].\] With a
non-smooth index function, two research questions are raised:

\begin{enumerate}
\def\labelenumi{\arabic{enumi})}
\tightlist
\item
  whether any optimiser fails to optimise this non-smooth index; and
\item
  whether the optimisers can find the global optimum despite the
  presence of a local optimum
\end{enumerate}

\begin{Schunk}
\begin{figure}

{\centering \includegraphics[width=1\linewidth]{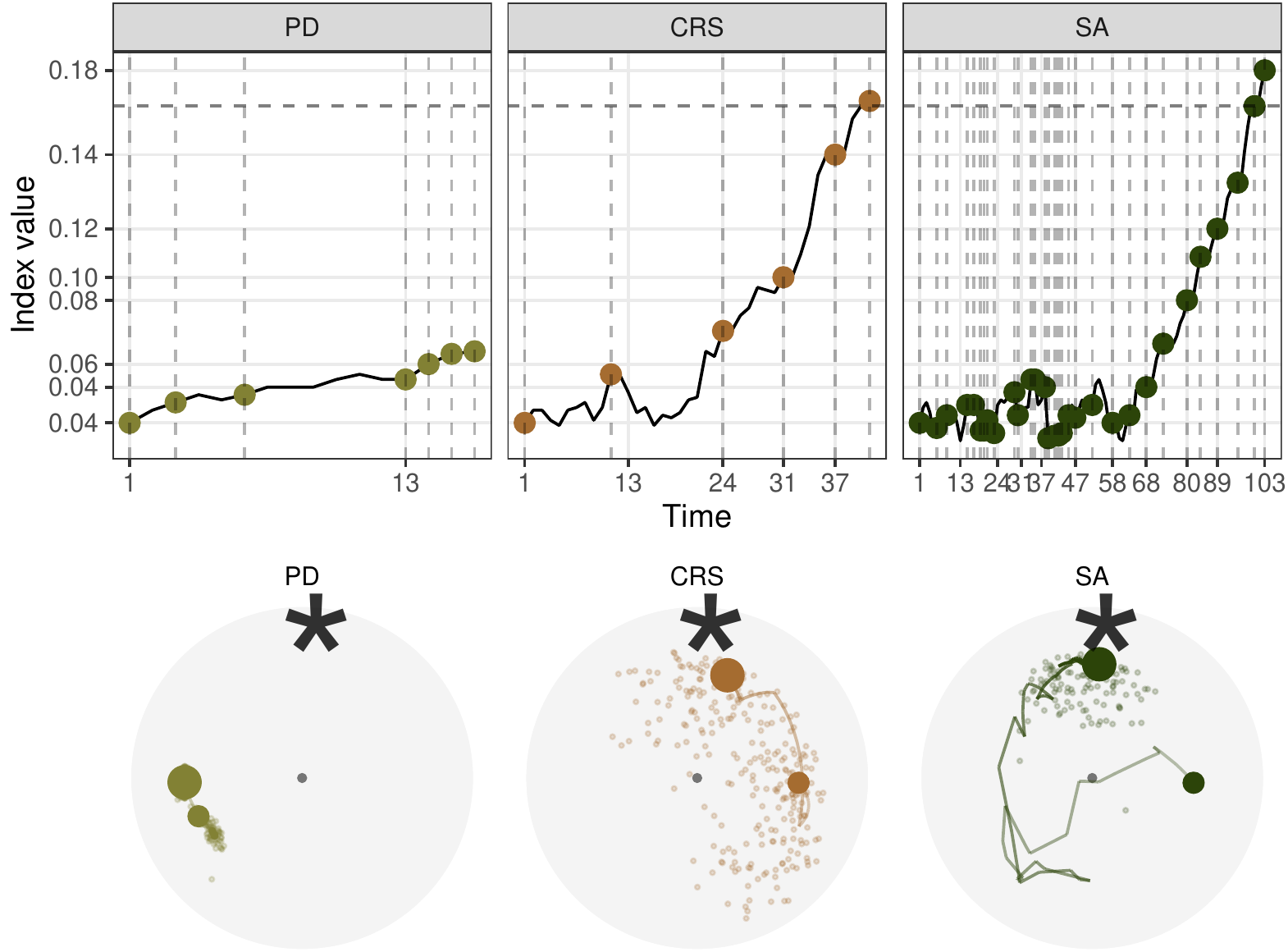}

}

\caption[Comparison of the three optimisers in optimising $I^{nk}(n)$ index for a 1D projection problem on a five-variable dataset, \code{boa5}]{Comparison of the three optimisers in optimising $I^{nk}(n)$ index for a 1D projection problem on a five-variable dataset, \code{boa5}. Both CRS and SA succeed in the optimisation, PD fails to optimise this non-smooth index. Further, SA takes  much longer than CRS to finish the optimisation, but finishes off closer to the theoretical best.}\label{fig:noisy-better-geo}
\end{figure}
\end{Schunk}

Figure \ref{fig:noisy-better-geo} presents the trace and PCA plots of
all three optimisers and as expected, none of the interpolated paths are
smooth. There is barely any improvement made by PD, indicating its
failure in optimising non-smooth index functions. While CRS and SA have
managed to get close to the index value of the theoretical best, the
trace plot shows that it takes SA much longer to find the final basis.
This long interpolation path is partially due to the fluctuation in the
early iterations, where SA tends to generously accept inferior bases
before concentrating near the optimum. The PCA plot shows the
interpolation path and search points excluding the last termination
iteration. Pseudo-Derivative (PD) quickly gets stuck near the starting
position. Comparing the amount of random search done by CRS and SA, it
is surprising that SA does not carry as many samples as CRS. Combining
the insights from both the trace and PCA plot, one can learn the two
different search strategies by CRS and SA: CRS frequently samples in the
space and only make a move when an improvement is guaranteed to be made,
while SA first broadly accepts bases in the space and then starts the
extensive sampling in a narrowed subspace. The specific decision of
which optimiser to use will depend on the index curve in the basis space
but if the basis space is non-smooth, accepting inferior bases at first,
as SA has done here, can lead to a more efficient search, in terms of
the overall number of points evaluated.

\begin{Schunk}
\begin{figure}

{\centering \includegraphics[width=1\linewidth]{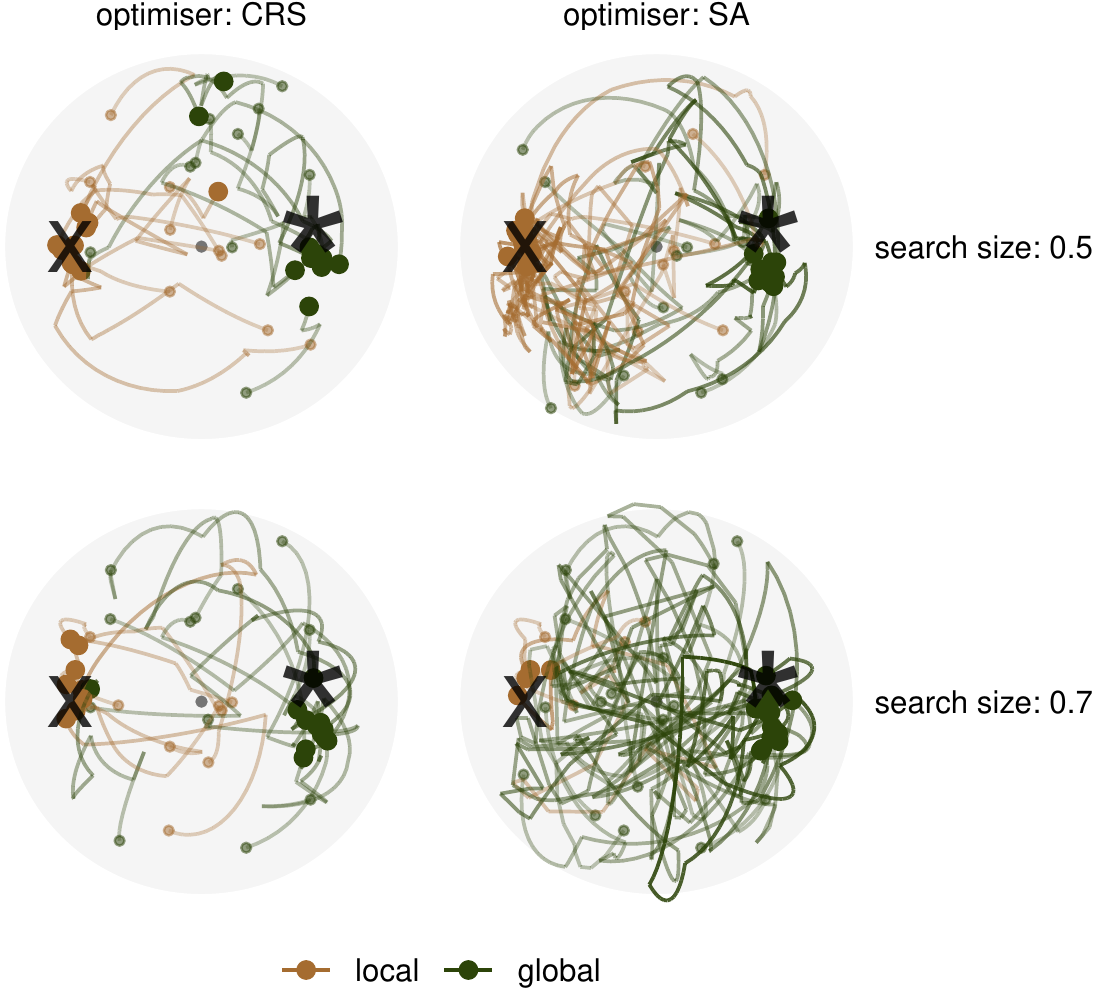}

}

\caption[Comparing 20 search paths in the PCA-projected basis space faceted by two optimisers]{Comparing 20 search paths in the PCA-projected basis space faceted by two optimisers: CRS and SA, and two search sizes: 0.5 and 0.7. The optimisation is on the 1D projection index, $I^{nk}(n)$, for \code{boa6} data, where a local optimum, annotated by the cross (x), is presented in this experiment, along with the global optimum (*).}\label{fig:kol-result}
\end{figure}
\end{Schunk}

The next experiment compares the performance of CRS and SA when a local
maximum exists. Two search neighbourhood sizes: 0.5 and 0.7 are compared
to understand how a large search neighbourhood would affect the final
basis and the length of the search. Figure \ref{fig:kol-result} shows 80
paths simulated using 20 seeds in the PCA plot, faceted by the optimiser
and search size. With CRS and a search size of 0.5, despite being the
simplest and fastest, the optimiser fails in three instances where the
final basis lands neither near the local nor the global optimum. With a
larger search size of 0.7, more seeds have found the global maximum.
Comparing CRS and SA for a search size of 0.5, SA does not seem to
improve the final basis found, despite having longer interpolation
paths. However, the denser paths near the local maximum is an indicator
that SA is working hard to examine if there is any other optimum in the
basis space but the relatively small search size has diminished its
ability to reach the global maximum. With a larger search size, almost
all the seeds (16 out of 20) have found the global maximum and some
final bases are much closer to the theoretical best, as compared to the
three other cases. This indicates that SA, with a reasonable large
search window, is able to overcome the local optimum and optimise close
towards the global optimum.

\hypertarget{reconciling-the-orientation}{%
\subsection{Reconciling the
orientation}\label{reconciling-the-orientation}}

\begin{Schunk}
\begin{figure}

{\centering \includegraphics[width=1\linewidth]{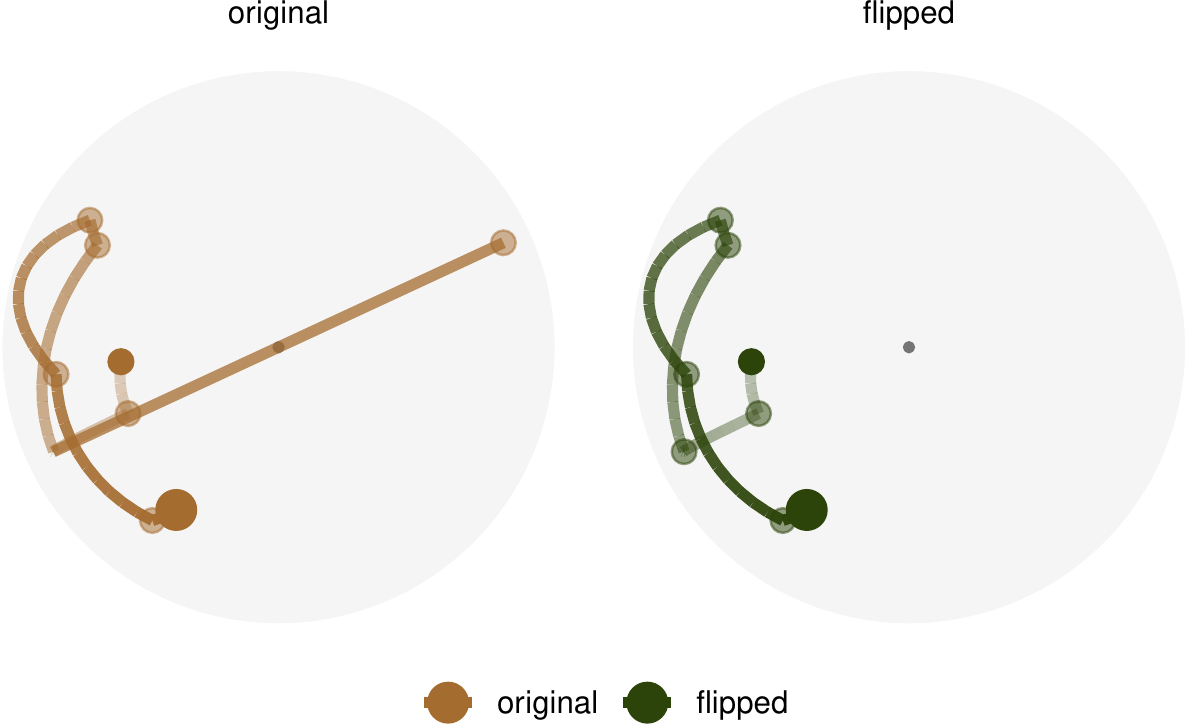}

}

\caption[Comparison of the interpolation in the PCA-projected basis space before and after reconciling the orientation of the target basis]{Comparison of the interpolation in the PCA-projected basis space before and after reconciling the orientation of the target basis. Optimisation is on the 1D projection index, $I^{nk}(n)$, for boa6 data using CRS with seed 2463. The dots represent the target basis in each iteration and the path shows the interpolation. On the left panel, one target basis is generated with an opposite orientation to the current basis (hence appear on the other side of the basis space) and the interpolator crosses the origin to perform the interpolation. The right panel shows the same interpolation after implementing an orientation check and the undesirable interpolation disappears.}\label{fig:flip-sign}
\end{figure}
\end{Schunk}

One interesting situation observed in the previous examples is that, for
some simulations, as shown on the left panel of Figure
\ref{fig:flip-sign}, the target basis is generated on the other half of
the basis space and the interpolator seems to draw a straight line to
interpolate. As mentioned previously, bases with opposite signs do not
affect the projection and index value, but clearly, we prefer the target
to have the same orientation as the current basis. The orientation of
two bases can be checked via calculating the determinant and a negative
value indicates opposite direction. Hence an orientation check is
carried out once a new target basis is generated and the sign in the
target basis will be flipped if a negative determinant is obtained. The
interpolation after implementing the orientation check is shown on the
right panel of Figure \ref{fig:flip-sign} where the unsatisfactory
interpolation no longer exists.

\hypertarget{implementation}{%
\section{Implementation}\label{implementation}}

This project contributes to the software development in two packages: a
data collection object is implemented in \pkg{tourr} \citep{tourr},
while the visual diagnostics of the optimisers is implemented in
\pkg{ferrn} \citep{ferrn}. The functions in the \pkg{ferrn}
\citep{ferrn} package are listed below:

\begin{itemize}
\item
  Main plotting functions:

  \begin{itemize}
  \tightlist
  \item
    \code{explore\_trace\_search()} produces summary plots in Figure
    \ref{fig:toy-search}.
  \item
    \code{explore\_trace\_interp()} produces trace plots for the
    interpolation points in Figure \ref{fig:toy-interp}.
  \item
    \code{explore\_space\_pca()} produces the PCA plot of projection
    bases on the reduced space in Figure \ref{fig:toy-pca}. The animated
    version in Figure \ref{fig:toy-pca-animated} can be turned on via
    the argument \texttt{animate\ =\ TRUE}.
  \item
    \code{explore\_space\_tour()} produces animated tour view on the
    full space of the projection bases in Figure \ref{fig:toy-tour}.
  \end{itemize}
\item
  \code{get\_*()} extracts and manipulates certain components from the
  existing data object.

  \begin{itemize}
  \tightlist
  \item
    \code{get\_anchor()} extracts target observations
  \item
    \code{get\_basis\_matrix()} flattens all the bases into a matrix
  \item
    \code{get\_best()} extracts the observation with the highest index
    value in the data object
  \item
    \code{get\_dir\_search()} extracts directional search observations
    for PD search
  \item
    \code{get\_interp()} extracts interpolated observations
  \item
    \code{get\_interp\_last()} extracts the ending interpolated
    observations in each iteration
  \item
    \code{get\_interrupt()} extracts the ending interpolated
    observations and the target observations if the interpolation is
    interrupted
  \item
    \code{get\_search()} extracts search observations
  \item
    \code{get\_search\_count()} extracts the count of search
    observations
  \item
    \code{get\_space\_param()} produces the coordinates of the centre
    and radius of the basis space
  \item
    \code{get\_start()} extracts the starting observation
  \item
    \code{get\_theo()} extracts the theoretical best observations, if
    given
  \end{itemize}
\item
  \code{bind\_*()} incorporates additional information outside the tour
  optimisation into the data object.

  \begin{itemize}
  \tightlist
  \item
    \code{bind\_theoretical()} binds the theoretical best observation in
    simulated experiment
  \item
    \code{bind\_random()} binds randomly generated bases in the
    projection bases space to the data object
  \item
    \code{bind\_random\_matrix()} binds randomly generated bases and
    outputs in a matrix format
  \end{itemize}
\item
  \code{add\_*()} provides wrapper functions to create ggprotos for
  different components for the PCA plot

  \begin{itemize}
  \tightlist
  \item
    \code{add\_anchor()} for plotting anchor bases
  \item
    \code{add\_anno()} for annotating the symmetry of start bases
  \item
    \code{add\_dir\_search()} for plotting the directional search bases
    with magnified distance
  \item
    \code{add\_end()} for plotting end bases
  \item
    \code{add\_interp()} for plotting the interpolation path
  \item
    \code{add\_interp\_last()} for plotting the last interpolation bases
    for comparing with target bases when interruption is used
  \item
    \code{add\_interrupt()} for linking the last interpolation bases
    with target ones when interruption is used
  \item
    \code{add\_search()} for plotting search bases
  \item
    \code{add\_space()} for plotting the circular space
  \item
    \code{add\_start()} for plotting start bases
  \item
    \code{add\_theo()} for plotting theoretical best bases, if
    applicable
  \end{itemize}
\item
  Utilities

  \begin{itemize}
  \tightlist
  \item
    \code{theme\_fern()} and \code{format\_label()} for better display
    of the grid lines and axis formatting
  \item
    \code{clean\_method()} to clean up the name of the optimisers
  \item
    \code{botanical\_palettes} is a collection of colour palettes from
    Australian native plants. Quantitative palettes include daisy,
    banksia, and cherry and sequential palettes contain fern and acacia
  \item
    \code{botanical\_pal()} as the colour interpolator
  \item
    \code{scale\_color\_*()} and \code{scale\_fill\_*()} for scaling the
    colour and fill of the plot
  \end{itemize}
\end{itemize}

\hypertarget{conclusion}{%
\section{Conclusion}\label{conclusion}}

This paper has provided several visual diagnostics that can be used for
diagnosing a complex optimisation procedure. The methods were
illustrated using the optimisers available for projection pursuit guided
tour. Here the constraint is the orthonormality condition of the
projection bases. The approach used broadly applies to other constrained
optimisers.

This work might be considered an effort to bring transparency into
algorithms. Although little attention is paid by algorithm developers to
providing ways to output information during intermediate steps, this is
an important component for enabling others to understand and diagnose
the performance. Algorithms are an essential component of artificial
intelligence that is used to make daily life easier. Interpretability of
algorithms is important to guard against aspects like bias and
inappropriate use. The data object underlying the visual diagnostics
here is an example of what might be useful in algorithm development
generally.

\hypertarget{acknowledgements}{%
\section{Acknowledgements}\label{acknowledgements}}

This article is created using \CRANpkg{knitr} \citep{knitr} and
\CRANpkg{rmarkdown} \citep{rmarkdown} in R. The source code for
reproducing this paper can be found at:
\url{https://github.com/huizezhang-sherry/paper-tour-vis}.

\bibliography{zhang-cook-laa-langrene-menendez}

\newpage

\hypertarget{appendix}{%
\section{Appendix}\label{appendix}}

Three algorithms (creeping random search, simulated annealing, and
pseudo-derivative) used in projection pursuit guided tour optimisation.

\begin{algorithm}
\SetAlgoLined
  \SetKwInOut{input}{input}
  \SetKwInOut{output}{output}
    \input{$f(.)$, $\alpha_1$, $l_{\max}$, $\text{cooling}$}
    \output{$\mathbf{A}_{l}$}
    Generate random start $\mathbf{A}_1$ and set $\mathbf{A}_{\text{cur}} \coloneqq \mathbf{A}_1$, $I_{\text{cur}} = f(\mathbf{A}_{\text{cur}})$, $j = 1$\;
  \Repeat{$\mathbf{A}_l$ is too close to $\mathbf{A}_{\text{cur}}$ in terms of geodesic distance}{
   Set $l = 1$\;
  \Repeat{$l > l_{\max}$ or $I_{l} > I_{\text{cur}}$}{
    Generate $\mathbf{A}_{l} = (1- \alpha_j)\mathbf{A}_{\text{cur}} + \alpha_j \mathbf{A}_{\text{rand}}$ and orthogonalise $\mathbf{A}_{l}$\;
    Compute $I_{l}  = f(\mathbf{A}_{l})$\;
    Update $l = l + 1$\;
  }
  Update $\alpha_{j+1} = \alpha_j * \text{cooling}$\;
  Construct the geodesic interpolation between $\mathbf{A}_{\text{cur}}$ and $\mathbf{A}_l$\;
  Update $\mathbf{A}_{\text{cur}} = \mathbf{A}_l$ and $j = j + 1$\;
}
  \caption{Creeping random search (CRS)}
  \label{random-search}
\end{algorithm}

\begin{algorithm}
\SetAlgoLined
\Repeat{$l > l_{\max}$ or $I_{l} > I_{\text{cur}}$ or $P > U$}{
    Generate $\mathbf{A}_{l} = (1- \alpha_j)\mathbf{A}_{\text{cur}} + \alpha_j \mathbf{A}_{\text{rand}}$ and orthogonalise $\mathbf{A}_{l}$\;
    Compute $I_{l}  = f(\mathbf{A}_{l})$, $T(l) = \frac{T_0}{\log(l + 1)}$ and $P= \min\left\{\exp\left[-\frac{I_{\text{cur}} -I_{l}}{T(l)}\right],1\right\}$\;
    Draw $U$ from a uniform distribution: $U \sim \text{Unif(0, 1)}$\;
    Update $l = l + 1$\;
  }
  \caption{Simulated annealing (SA)}
  \label{simulated-annealing}
\end{algorithm}

\begin{algorithm}
\SetAlgoLined
\Repeat{$l > l_{\max}$ or $p_{\text{diff}} > 0.001$}{
  Generate $n$ random directions $\mathbf{A}_{\text{rand}}$ \;
  Compute $2n$ candidate bases deviate from $\mathbf{A}_{\text{cur}}$ by an angle of $\delta$ while ensuring orthogonality\;
  Compute the corresponding index value for each candidate bases\;
  Determine the search direction as from $\mathbf{A}_{\text{cur}}$ to the candidate bases with the largest index value\;
  Determine the step size via optimising the index value on the search direction over a 90 degree window\;
  Find the optimum $\mathbf{A}_{**}$ and compute $I_{**} = f(\mathbf{A}_{**})$, $p_{\text{diff}} = (I_{**} - I_{\text{cur}})/I_{**}$\;
  Update $l = l + 1$\;
}
\caption{Pseudo-derivative (PD)}
\label{search-geodesic}
\end{algorithm}

\address{%
H.Sherry Zhang\\
Monash University\\%
Department of Econometrics and Business Statistics\\
\\\href{mailto:huize.zhang@monash.edu}{\nolinkurl{huize.zhang@monash.edu}}
}

\address{%
Dianne Cook\\
Monash University\\%
Department of Econometrics and Business Statistics\\
Melbourne, Australia%
\\\href{mailto:dicook@monash.edu}{\nolinkurl{dicook@monash.edu}}
}

\address{%
Ursula Laa\\
University of Natural Resources and Life Sciences\\%
Institute of Statistics\\
Vienna, Austria%
\\\href{mailto:ursula.laa@boku.ac.at}{\nolinkurl{ursula.laa@boku.ac.at}}
}

\address{%
Nicolas Langrené\\
CSIRO Data61\\%
34 Village Street, Docklands VIC 3008 Australia\\
Melbourne, Australia%
\\\href{mailto:nicolas.langrene@csiro.au}{\nolinkurl{nicolas.langrene@csiro.au}}
}

\address{%
Patricia Menéndez\\
Monash University\\%
Department of Econometrics and Business Statistics\\
Melbourne, Australia%
\\\href{mailto:patricia.menendez@monash.edu}{\nolinkurl{patricia.menendez@monash.edu}}
}
\end{article}

\end{document}